\documentclass[fleqn,twoside,11pt]{article}

\usepackage{amssymb}
\usepackage{graphicx}
\usepackage{amsmath}

\begin{document}



\begin{center}
{\Large\bf Charged Lepton Mass Formula \footnote{
Talk at Internationa Workshop on Neutrino Masses and Mixing
--- Toward Unified Understanding of Quark and Lepton Mass Matrices --- ,
at Shizuoka, Japan, December, 17-19, 2006;
To be published in IJMP E, July (2007)} \\
-- Development and Prospect -- \\
}

{\footnotesize Yoshio Koide 
\footnote{Present address: Department of Physics, Osaka University,
Toyonaka, Osaka, 560-0043, Japan\\
E-mail address: koide@het.phys.sci.osaka-u.ac.jp}
}

{\it Department of Physics, University of Shizuoka,\\
 52-1 Yada, 
Suruga-ku, Shizuoka-shi, 422-8526,
Japan\\
koide@u-shizuoka-ken.ac.jp}

\end{center}





\begin{abstract}
The recent devolopment on the charged lepton mass forumula
$m_e+m_{\mu}+m_{\tau}=
\frac{2}{3}\left( \sqrt{m_e}+\sqrt{m_\mu}\right.$ 
$\left. +\sqrt{m_{\tau}}\right)^2$ is reviewed.
An S$_3$ or A$_4$ model will be promising for the mass
relation.
\end{abstract}

\section{Beginning of the charged lepton mass formula}

It is generally considered that masses and mixings  of the quarks 
and leptons will obey a simple law of  nature, so that 
we expect that we will find a beautiful relation among those 
values.  It is also considered that the mass 
matrices of the fundamental particles will be 
governed by a symmetry.

Our dream is to understand 
masses and mixings from a symmetry, i.e. 
not from adjustable parameters,
but from Clebsch-Gordan-like 
coefficients.

For example, in 1971, the author\cite{Koide_Cab} 
tried to understand the Cabibbo mixing from a U(3) 
symmetry from a composite model of quarks, 
the so-called Hiroshima model\cite{Hiroshima}.
\footnote{In the model, the quarks are composed as
$$
\begin{array}{l}
u=\langle (\nu_e \overline{\nu}_e \cos\theta +
\nu_\mu \overline{\nu}_\mu \sin\theta \rangle)B
\rangle , \\
c=\langle (-\nu_e \overline{\nu}_e \sin\theta +
 \nu_\mu \overline{\nu}_\mu \cos\theta \rangle)B
\rangle , \\
d= \langle (e^- \overline{\nu}_e)B \rangle , \\
s=  \langle (\mu^- \overline{\nu}_\mu)B \rangle ,
\end{array}
$$
where $B$ is a hypothetical fermion with the baryon number.
In the model, semileptonic decays $ d \rightarrow u$ and 
$ s \rightarrow u$ are understood from $ e^-\rightarrow \nu_e$ 
with the factor $\cos\theta$ and from $\mu^- \rightarrow \nu_\mu$ 
with the factor $\sin\theta$, respectively.
The nonleptonic decay $s \rightarrow d$ can be understood from
 $\mu^- \overline{\nu}_\mu \rightarrow e^- \overline{\nu}_e$ 
with the factor 1 without assuming the so-called penguin diagram.}
The basic idea\cite{Koide_Cab} to give the Cabibbo angle was 
analogy to the hadronic $\pi^0$-$\eta^0$-$\sigma^0$ mixing 
where we assume 
$(\nu_e, \nu_\mu, \ell) \sim {\bf 3}$ of U(3),
and 3 quarks are members of {\bf 8+1} of U(3).
In the model, when we define the following lepton-antilepton states
$$
\begin{array}{l}
\pi =\frac{1}{\sqrt{2}}(\nu_e \overline{\nu}_e -\nu_\mu \overline{\nu}_\mu), \\
\eta=\frac{1}{\sqrt{6}}(\nu_e \overline{\nu}_e +\nu_\mu \overline{\nu}_\mu
-2 \ell \overline{\ell}), \\
\sigma=\frac{1}{\sqrt{3}}(\nu_e \overline{\nu}_e +\nu_\mu \overline{\nu}_\mu
+\ell \overline{\ell}),
\end{array}
\eqno(1.1)
$$
$$
\begin{array}{l}
\pi =\frac{1}{\sqrt{2}}(\nu_e \overline{\nu}_e -\nu_\mu \overline{\nu}_\mu), \\
\omega=\frac{1}{\sqrt{2}}(\nu_e \overline{\nu}_e
+\nu_\mu \overline{\nu}_\mu), \\
\phi=\frac{1}{\sqrt{3}}\ell \overline{\ell},
\end{array}
\eqno(1.2)
$$
from the analogy with the hadoronic $\pi$-$\eta$-$\eta'$ mixing, 
we consider that the physical up-quark $u$ has the lepton-antilepton state
$$
\pi' =\frac{\sqrt{3}}{2} \pi +\frac{1}{2} \omega
=\frac{\sqrt{3}+1}{2\sqrt{2}} \nu_e \overline{\nu}_e -
\frac{\sqrt{3}-1}{2\sqrt{2}} \nu_\mu \overline{\nu}_\mu,
\eqno(1.3)
$$ 
so that the Cabibbo angle $\theta_C$ is given by
$$
\sin\theta_C=\frac{\sqrt{3}-1}{2 \sqrt{2}}, \ \ \ 
\cos\theta_C=\frac{\sqrt{3}+1}{2 \sqrt{2}}.
\eqno(1.4)
$$
However, the derivation of (1.4) was  strained and tricky. 
In 1978, Harari {\it et al.}\cite{Harari} have proposed a model based 
on a permutation symmetry S$_3$ and
they have successfully derived the relation (1.4) for the Cabibbo angle
$\theta_C$.

In 1982, the author\cite{Koidemass} have proposed 
a charged lepton mass relation\footnote{
Exactly speaking, the expression (1.5) appeared in Ref.~\cite{Koide90}.
In Refs.\cite{Koidemass}, the formula has been given by the expression
$m_{ei}=m_0 (x_i + x_0)^2$, where $x_1+x_2+x_3=0$ and 
$x_0=\sqrt{(x_1^2+x_2^2+x_3^2)/3}$.
}
$$
m_e+m_{\mu}+m_{\tau}=
\frac{2}{3}\left( \sqrt{m_e}+\sqrt{m_\mu}+\sqrt{m_{\tau}} 
\right)^2 .
\eqno(1.5)
$$
It is well-known that the observed charged lepton mass spectrum satisfies 
the relation with remarkable precision. 

The mass formula (1.5) is invariant under any exchange 
$\sqrt{m_i}\leftrightarrow\sqrt{m_j}$. 
This suggests that a description by a permutation 
symmetry\cite{S3} S$_3$ may be useful 
for the mass matrix model. 
However, this formula was derived from an extension of the 
$\pi$-$\eta$-$\sigma$ mixing model (composite model), 
and the author had never considered an S$_3$ symmetry although he had still 
assumed a U(3) symmetry until 1999\cite{Koide99}.

In the present paper, we will demonstrate how the S$_3$ symmetry is 
promising to understand the masses and mixings.   
And, we will comment that an A$_4$ symmetry is also promising.
Although we think that there are beautiful relations among quark and 
lepton masses and mixings, it is hard to see such a symmetry in 
the quark sector, because the original symmetry will be spoiled by 
the gluon cloud. 
In the present study, we will confine ourselves to the investigation 
of the lepton masses and mixings.

\section{Leptons under S$_3$}

 Let us demonstrate that the S$_3$ symmetry is promising not only 
for understanding of the charged lepton mass relation (1.5), 
but also for that of the observed neutrino mixing, i.e. 
the so-called tribimaximal mixing\cite{tribi}
$$
U_{TB}=\left(\begin{array}{ccc}
-\frac{2}{\sqrt6} & \frac{1}{\sqrt3} & 0 \\
\frac{1}{\sqrt6} & \frac{1}{\sqrt3} & -\frac{1}{\sqrt2} \\
\frac{1}{\sqrt6} & \frac{1}{\sqrt3} & \frac{1}{\sqrt2} \\
\end{array} \right) .
\eqno(2.1)
$$

\subsection{Tribimaximal mixing}

We define the doublet $(\psi_\pi,\psi_\eta)$ and singlet $\psi_\sigma$ 
of S$_3$
$$
\left(\begin{array}{l}
\psi_\pi \\
\psi_\eta \\
\psi_\sigma \\
\end{array} \right)=
\left(\begin{array}{ccc}
0 & -\frac{1}{\sqrt2} & \frac{1}{\sqrt2} \\
-\frac{2}{\sqrt6} & \frac{1}{\sqrt6} & \frac{1}{\sqrt6} \\
\frac{1}{\sqrt3} & \frac{1}{\sqrt3} & \frac{1}{\sqrt3} \\
\end{array} \right)
\left(\begin{array}{l}
\psi_1 \\
\psi_2 \\
\psi_3 \\
\end{array} \right).
\eqno(2.2)
$$
If the neutrino mass eigenstates are $(\nu_\pi,\nu_\eta,\nu_\sigma)$ 
defined by the relation (2.2) 
with the mass hierarchy $m_{\nu\eta}^2<m_{\nu\sigma}^2<m_{\nu\pi}^2$ 
in contrast to the weak eigenstates 
$(\nu_1,\nu_2,\nu_3)=(\nu_e,\nu_\mu,\nu_\tau)$,
then we can obtain the tribimaximal mixing
(2.1) because of the relation
$$
\left(\begin{array}{l}
\nu_e \\
\nu_\mu \\
\nu_\tau \\
\end{array} \right)=
\left(\begin{array}{l}
\nu_1 \\
\nu_2 \\
\nu_3 \\
\end{array} \right)=
\left(\begin{array}{ccc}
-\frac{2}{\sqrt6} & \frac{1}{\sqrt3} & 0 \\
\frac{1}{\sqrt6} & \frac{1}{\sqrt3} & -\frac{1}{\sqrt2} \\
\frac{1}{\sqrt6} & \frac{1}{\sqrt3} & \frac{1}{\sqrt2} \\
\end{array} \right)
\left(\begin{array}{l}
\nu_\eta \\
\nu_\sigma \\
\nu_\pi \\
\end{array} \right).
\eqno(2.3)
$$

\subsection{Charged lepton mass formula and Higgs potential}

The mass formula (1.5) can be understood
 from a universal seesaw model\cite{Koide90} with 3-flavor
 scalars $\phi_i$: 
$$
M_f=m_L^fM_F^{-1}m_R^f .
\eqno(2.4)
$$
Here, for the charged lepton sector, we take  
$$
m_L^e=\frac{1}{\kappa}m_R^e=y_e{\rm diag}(v_1,v_2,v_3), 
\eqno(2.5)
$$
where the VEV $v_i=\langle \phi_i \rangle$ satisfy the relation
$$
v_1^2 + v_2^2 + v_3^2 = \frac{2}{3}
 \left( v_1 + v_2 + v_3 \right) ^2 .
\eqno(2.6)
$$ 
The VEV relation (2.6) means
$$
 v_\pi^2+v_\eta^2 =v_\sigma^2 ,
\eqno(2.7)
$$
because 
$$
v_1^2+v_2^2+v_3^2=v_\pi^2+v_\eta^2+v_\sigma^2 
= 2 v_\sigma^2
=2 \left( \frac{v_1+v_2+v_3}{\sqrt{3}} \right)^2.
\eqno(2.8)
$$

The Higgs potential which gives the relation (2.7) 
is,  for example, given by\cite{Koide06}
$$
V=\mu^2 \sum_{i}(\overline{\phi}_i \phi_i)
+\frac{1}{2} \lambda_1 \left[\sum_i(\overline{\phi}_i \phi_i)\right]^2
$$
$$
+ \lambda_2 (\overline{\phi}_{\sigma} \phi_{\sigma})
(\overline{\phi}_{\pi}\phi_{\pi} + \overline{\phi}_{\eta}\phi_{\eta})
+V_{SB} ,
\eqno(2.9)
$$
where $V_{SB}$ is a soft symmetry breaking term which does 
not affect the derivation of (2.7).
Here, the existence of the S$_3$ invariant $\lambda_2$-term 
is essential for the derivation of the VEV relation (2.7). 
Note that the relation (2.7)  can
be obtained independently of the explicit values of
the parameter $\lambda_2$. 

\subsection{Yukawa interaction form under S$_3$ }

The general form of the S$_3$ invariant Yukawa interaction is
 given by

$$
H=\left(
y_0\frac{\bar{\psi}_\pi \psi_\pi+\bar{\psi}_\eta \psi_\eta+\bar{\psi}_\sigma 
\psi_\sigma}{\sqrt3}+y_1\frac{\bar{\psi}_\pi \psi_\pi+\bar{\psi}_\eta 
\psi_\eta-2\bar{\psi}_\sigma \psi_\sigma}{\sqrt6}
\right) \phi_\sigma 
$$
$$
+y_2 \left(
\frac{\bar{\psi}_\pi \psi_\eta+\bar{\psi}_\eta \psi_\pi}{\sqrt2}\phi_\pi
+\frac{\bar{\psi}_\pi \psi_\pi-\bar{\psi}_\eta \psi_\eta}{\sqrt2}\phi_\eta
\right)
$$
$$
+y_3\frac{\bar{\psi}_\pi \phi_\pi+\bar{\psi}_\eta \phi_\eta}{\sqrt2}\psi_\sigma
+y_4\bar{\psi}_\sigma \frac{\phi_\pi \psi_\pi+\phi_\eta \psi_\eta}{\sqrt2}.
\eqno(2.10)
$$

For the charged lepton sector, we have already
 assumed the form 
$$
H_e = y_e \left( \bar{\ell}_{L1} \phi_{L1}^d E_{R1} 
+\bar{\ell}_{L2} \phi_{L2}^d E_{R2}+
\bar{\ell}_{L3}\phi_{L3}^d E_{R3} \right).
\eqno(2.11)
$$ 
The form (2.11) corresponds to the case 
$$
y_0=y_e, \ \ y_1=0, \ \ y_2=\frac{1}{\sqrt3}y_e, \ \ y_3=y_4=
\sqrt{\frac{2}{3}}y_e ,
\eqno(2.12)
$$
in the general form (2.10).

The general study under the S$_3$ symmetry will be found 
in  a recent paper\cite{Koide0605} by the author. 
Here, the details are skiped.

\subsection{A simple example of the S$_3$-invariant neutrino interaction}

Now, let us speculate an S$_3$-invariant neutrino
 interaction with a concise form\cite{Koide0603} 
$$
H_\nu=y_\nu \left( \frac{\overline{\ell}_\pi N_\pi
+\overline{\ell}_\eta N_\eta
+\overline{\ell}_\sigma N_\sigma}{\sqrt3}\phi_\sigma^u
+ \frac{\overline{\ell}_\pi N_\eta
+\overline{\ell}_\eta N_\pi}{\sqrt2}\phi_\pi^u
+\frac{\overline{\ell}_\pi N_\pi
-\overline{\ell}_\eta N_\eta}{\sqrt2}\phi_\eta^u \right) .
\eqno(2.13)
$$
Here, we have assumed
the universality of the coupling constants 
to the $\phi_\sigma$, $\phi_\pi$ and $\phi_\eta$ terms. 
Then, we obtain a simple mass spectrum
$$
\begin{array}{lll}
m_{\nu 1}=\left(\frac{1}{\sqrt6}-\frac{1}{2} \right)^2 m_0^\nu , \\
m_{\nu 2}=\frac{1}{6}m_0^\nu , \\
m_{\nu 3}=\left(\frac{1}{\sqrt6}+\frac{1}{2} \right)^2 m_0^\nu , \\
\end{array}
\eqno(2.14)
$$
where we have used the relation 
$$
v_\pi^2+v_\eta^2=v_\sigma^2 ,
\eqno(2.15)
$$
from the S$_3$ Higgs potential model.
By putting $m_{\nu 3} = \sqrt{\Delta m^2_{atm}}$ from the atmospheric
neutrino oscillation data\cite{atm}, we obtain 
$$
\begin{array}{l}
m_{\nu 1}=(5.3_{-0.3}^{+0.4})\times10^{-4}\ {\rm eV}, \\
m_{\nu 2}=(1.05_{-0.05}^{+0.07})\times10^{-2}\ {\rm eV}, \\
m_{\nu 3}=(5.22_{-0.25}^{+0.35})\times10^{-2}\ {\rm eV}.
\end{array}
\eqno(2.16)
$$

The present case predicts
$$
R=\frac{\Delta m_{21}^2}{\Delta m_{32}^2}=\frac{4\sqrt6-9}{4\sqrt6+9}=0.0423.
\eqno(2.17)
$$
The predicted value is somewhat larger than the observed value\cite{solar,atm}
$$
R_{obs} \equiv \frac{\Delta m^2_{solar}}{\Delta m^2_{atm}}
=\frac{(7.9_{-0.5}^{+0.6})\times10^{-5}{\rm eV}^2}{
(2.72_{-0.25}^{+0.38})\times10^{-3}{\rm eV}^2}.
=(2.9\pm0.5)\times10^{-2}
\eqno(2.18)
$$
However, the case cannot, at present, be  ruled 
out within three sigma.

 If $v_\pi/v_\eta \neq 0$ , $m_L^\nu$ cannot be diagonal on the
 basis $(\nu_\pi,\nu_\eta,\nu_\sigma)$. 
The mixing angle $\theta_{\pi\eta}$ of the further rotation between 
$\nu_\pi$-$\nu_\eta$ is given by 
$$
\tan \theta_{\pi\eta} = v_\pi/v_\eta.
\eqno(2.19)
$$ 

It is well known that the $2\leftrightarrow 3$ symmetry\cite{23sym} is
 promising for neutrino mass matrix description.
We also assume the $2\leftrightarrow 3$ symmetry for 
$\langle \phi^u_i\rangle$, i.e. $v_2^u=v_3^u$, which leads to 
$$
\langle \phi_\pi^u\rangle =0 .
\eqno(2.20)
$$ 
Therefore, the present model gives the exact 
tribimaximal mixing:
$$
\sin^2 2\theta_{23}=1 ,\ \ \tan^2\theta_{12}=1/2, \ \ 
\theta_{13}=0 .
\eqno(2.21)
$$
Note that if we require the $2\leftrightarrow 3$ symmetry for
 the fields $\ell_{Li}=(\nu_{Li},e_{Li})$, the symmetry
will affect the charged lepton sector, too.
Here, we have assumed the $2\leftrightarrow 3$ symmetry
only for $\langle \phi^u_i\rangle$, not for $\langle \phi^d_i\rangle$, 
so that the symmetry does not affect the charged lepton mass matrix.

\subsection{Summery of the S$_3$ model}

In conclusion, in the  S$_3$ model, the following assumptions
have been done:

\noindent (i) 
We have assumed a universal seesaw model.

\noindent (ii) 
The S$_3$ symmetry in the Yukawa interactions is strictly unbroken. 
The symmetry S$_3$ is broken only though the VEV of Higgs scalars $\phi_i$.

\noindent (iii) 
We have required the universality of the coupling constants on the basis 
$(e_1, e_2, e_3)$ for the charged lepton sector, 
while we have assumed that on the basis $(\nu_\pi,\nu_\eta, \nu_\sigma)$ 
for the neutrino sector.

As a result of those assumptions, we have obtainde the tribimaximal mixing
(2.1) and the charged lepton mass formula (1.5).



\section{Prospect of the Charged Lepton Mass Formula}

\subsection{Brannen's speculations}

Recently, Brannen has speculated a neutrino mass relation\cite{Brannen}
 similar to the charged lepton mass relation (1.5):
$$
m_{\nu 1}+m_{\nu 2}+m_{\nu 3}=
\frac{2}{3}\left( -\sqrt{m_{\nu 1}}+\sqrt{m_{\nu 2}}+\sqrt{m_{\nu 3}} 
\right)^2 .
\eqno(3.1)
$$
Of course, we cannot extract the values of the neutrino mass ratios 
$m_{\nu 1}/m_{\nu 2}$ and $m_{\nu 2}/m_{\nu 3}$ from the neutrino 
oscillation data 
$\Delta m^2_{solar}$ and $\Delta m^2_{atm}$ 
unless we have more information on the neutrino masses, 
so that we cannot judge whether the observed neutrino masses satisfy 
the relation (3.1) or not.

Generally, the masses which satisfy the relations (1.5) and (3.1) 
can be expressed as a bilinear form
$$
m_{fi} = (z_{fi})^2 m_{f0} ,
\eqno(3.2)
$$ 
where
$$
\begin{array}{lll}
z_{f1}=\frac{1}{\sqrt6}-\frac{1}{\sqrt3}\sin\xi_f,  \\
z_{f2}=\frac{1}{\sqrt6}-\frac{1}{\sqrt3}\sin(\xi_f+\frac{2}{3}\pi) , \\
z_{f3}=\frac{1}{\sqrt6}-\frac{1}{\sqrt3}\sin(\xi_f+\frac{4}{3}\pi), 
\end{array}
\eqno(3.3)
$$  
$$
(z_{f1})^2+(z_{f2})^2+(z_{f3})^2=1.
\eqno(3.4)
$$
Then, Brannen has also speculated the relation\cite{Brannen}
$$
\xi_\nu=\xi_e+\frac{\pi}{12} .
\eqno(3.5)
$$ 
From the observed charged lepton mass values,
 we obtain
$$
\xi_e=\frac{\pi}{4}-\varepsilon=42.7324^\circ \ \ 
(\varepsilon=2.2676^\circ) .
\eqno(3.6)
$$ 
Therefore, the Brannen relation (3.5) gives
$$
\xi_\nu=57.7324^\circ ,
\eqno(3.7)
$$   
which predicts
$$
R=\frac{\Delta m^2_{21}}{\Delta m^2_{32}} = 0.0318 .
\eqno(3.8)
$$

The value (3.8) is in good agreement with 
the observed value $0.029\pm 0.005$, (2.17).
Therefore, the speculations by Brannen are favorable to 
the observed neutrino data.

We can understand Brannen's first relation (3.1) 
by assuming two different scalars i.e. $\phi^u \neq \phi^d$.
However, Brannen's second relation (3.5) is  hard 
to be derived from the conventional symmetries.
it is an open question that the relation (3.5) is 
accidental or not.\cite{Koide0605}

Besides, Brannen\cite{Brannen} and Rosen\cite{Rosen} 
 have speculated that the observed value  
$\xi_e=42.7324^\circ$ is given by the relation
$$
\xi_e =\frac{\pi}{6} +\frac{2}{9}.
\eqno(3.9)
$$     
Since the value $2/9$ means
$$
\frac{2}{9}\, {\rm rad} = 12.732395^\circ ,
\eqno(3.10)
$$
the speculation (3.9) is in excellent agreement with the
observed value $\xi_e=42.7324^\circ$ from the charged lepton
masses. This is an amazing coincidence.

However, at present, there is no reason for the relation (3.9).
Too adhering to this coincidence will again push the formula 
(1.5) into a mysterious world, so that it is not recommended
that we take the relation (3.9) seriously at present.

\subsection{From Seesaw to Frogatt-Nielsen}

The seesaw model with the three SU(2)$_L$ doublet scalars 
$\phi_i$ causes the 
flavor changing neutral current (FCNC) problem.
Therefore, the seesaw model may, for example, be translated into 
a Frogatt-Nielsen-type model:
$$
H_{eff}=y_e \overline{l}_L H_L^d \frac{\phi^d}{\Lambda_d}
\frac{\phi^d}{\Lambda_d}e_R
+y_\nu \overline{l}_L H_L^u \frac{\phi^u}{\Lambda_u} \nu_R+y_R 
\overline{\nu}_R 
{\Phi}\nu_R^\ast .
\eqno(3.11)
$$

The argument about flavor structure is essentially unchanged 
under the present model-changing.
However, the scenario for the symmetry 
breaking energy scale will be considerably
changed: 
The energy scale of $v_i=\langle \phi_i \rangle$ in the 
seesaw model is of the order of $10^2$ GeV, while that in
the Frogatt-Nielsen\cite{Frogatt} type model is of the 
order of $10^{19}$ GeV.
In the Frogatt-Nielsen-type model, the S$_3$-broken
structure of the effective Yukawa coupling 
constants is given  at the Planck mass scale.
However, this is no problem, because the formula
(1.1) is not so sensitive to the renormalization group
equation (RGE) effects as far as
 the lepton sector is concerned.\cite{evol_KoideF}

\subsection{From S$_3$ to A$_4$}

Recently, Ma\cite{Ma06}  has 
explained the observed tribimaximal mixing 
on the basis of A$_4$.
This suggests that the A$_4$ symmetry is also
promising as a symmetry of leptons.

\subsubsection{Irreducible representations of A$_4$}

When we denote {\bf 3} of A$_4$ as
$$
\overline{\psi} = \left(
\begin{array}{c}
\overline{\psi}_1 \\
\overline{\psi}_2 \\
\overline{\psi}_3 
\end{array} \right) \sim {\bf 3}, \ \ 
{\psi} = \left(
\begin{array}{c}
{\psi}_1 \\
{\psi}_2 \\
{\psi}_3 
\end{array} \right) \sim {\bf 3},
\eqno(3.12)
$$
we can make {\bf 1}, {\bf 1}', {\bf 1}'' from $3\times 3$ as follows:
$$
\begin{array}{l}
(\overline{\psi}\psi)_{\bf 1}=\frac{1}{\sqrt{3}} 
(\overline{\psi}_1\psi_1 +\overline{\psi}_2\psi_2
+\overline{\psi}_3\psi_3 ), \\
(\overline{\psi}\psi)_{\bf 1'}=\frac{1}{\sqrt{3}} 
(\overline{\psi}_1\psi_1 +\overline{\psi}_2\psi_2 \omega
+\overline{\psi}_3\psi_3 \omega^2), \\
(\overline{\psi}\psi)_{\bf 1"}=\frac{1}{\sqrt{3}} 
(\overline{\psi}_1\psi_1 +\overline{\psi}_2\psi_2 \omega^2
+\overline{\psi}_3\psi_3 \omega), 
\end{array}
\eqno(3.13)
$$
where
$$
\omega = e^{i\frac{2}{3}\pi} = \frac{-1+i\sqrt{3}}{2}.
\eqno(3.14)
$$ 

When we define
$$
\begin{array}{l}
(\overline{\psi}\psi)_{\sigma}=\frac{1}{\sqrt{3}} 
(\overline{\psi}_1\psi_1 +\overline{\psi}_2\psi_2
+\overline{\psi}_3\psi_3 ) , \\
(\overline{\psi}\psi)_{\eta}=\frac{1}{\sqrt{6}} 
(2\overline{\psi}_1\psi_1 -\overline{\psi}_2\psi_2 
-\overline{\psi}_3\psi_3 ) , \\
(\overline{\psi}\psi)_{\pi}=\frac{1}{\sqrt{2}} 
(\overline{\psi}_3\psi_3 -\overline{\psi}_2\psi_2 ) ,
\end{array}
\eqno(3.15)
$$
we can express (3.13) as
$$
\begin{array}{l}
(\overline{\psi}\psi)_{\bf 1}=(\overline{\psi}\psi)_\sigma , \\
 (\overline{\psi}\psi)_{\bf 1'}=\frac{1}{\sqrt{2}} 
\left[(\overline{\psi}\psi)_\eta -i(\overline{\psi}\psi)_\pi\right] , \\
(\overline{\psi}\psi)_{\bf 1"}=\frac{1}{\sqrt{2}} 
\left[(\overline{\psi}\psi)_\eta +i(\overline{\psi}\psi)_\pi\right] .
\end{array}
\eqno(3.16)
$$
Therefore, if we define $(\phi_\pi, \phi_\eta, \phi_\sigma)$  as
$$
\begin{array}{l}
\phi_{\bf 1} =\phi_\sigma ,\\
\phi_{\bf 1'}=\frac{1}{\sqrt{2}} (\phi_\eta - i\phi_\pi) , \\
\phi_{\bf 1"}=\frac{1}{\sqrt{2}} (\phi_\eta + i\phi_\pi) ,
\end{array}
\eqno(3.17)
$$
we can write an A$_4$ invariant Yukawa interaction 
as follows:
$$
\begin{array}{l}
(\overline{\psi} \psi)_{\bf 1} \phi_{\bf 1} 
+(\overline{\psi} \psi)_{\bf 1'} \phi_{\bf 1"} 
+(\overline{\psi} \psi)_{\bf 1"} \phi_{\bf 1'} \\
=(\overline{\psi} \psi)_{\sigma} \phi_{\sigma} 
+(\overline{\psi} \psi)_{\eta} \phi_{\eta} 
+(\overline{\psi} \psi)_{\pi} \phi_{\pi} .
\end{array} 
\eqno(3.18)
$$ 
Thus, we can translate the relations in S$_3$ into those in A$_4$.
The A$_4$ symmetry will be also promising in the symmetry of 
the leptons\cite{Koide-A4}.

\subsubsection{
Higgs potential which gives $ v_\pi^2+v_\eta^2 =v_\sigma^2 $}

The existence of the $\lambda_2$-term, 
$\phi_\sigma^2 (\phi_\pi^2+\phi_\eta^2)$,
in the S$_3$ Higgs potential model (2.9)
 was  essential for the derivation of the relation (2.7).
In an A$_4$ model, the $\lambda_2$-term can be composed 
as
$$
\phi_{\bf 1} \phi_{\bf 1} \left( \phi_{\bf 1'}\phi_{\bf 1"}
+\phi_{\bf 1"} \phi_{\bf 1'}\right) 
= \phi_\sigma^2 \left( \phi_\pi^2 +\phi_\eta^2 \right). 
\eqno(3.19)
$$
Thus, the charged lepton mass relation (1.5) 
can also be derived from the A$_4$ model.  

\subsection{From non-SUSY to SUSY}

Since the formula (1.5) is so exact, we need to find a condition 
which is protected against large corrections.
Ma\cite{Ma-priv} has recently proposed a supersymmetric 
S$_3$-invariant Higgs potential,  
because supersymmetry is unbroken even after 
the superfields acquire VEVs:
$$
W = \frac{1}{2} m \left( \phi_\pi^2 + \phi_\eta^2 + \phi^2_\sigma 
\right) +\frac{1}{3} \lambda \phi_\sigma^3 
+\lambda \phi_\sigma \left(\phi_\pi^2 + \phi_\eta^2\right) ,
\eqno(3.20)
$$
which also leads to the relation  $ v_\pi^2+v_\eta^2 =v_\sigma^2 $, (2.7).

By extending this idea, and by applying 
a symmetry $\Sigma(81)$ to the model, Ma\cite{Ma0612} has 
proposed  a new model which leads to the 
charged lepton mass relation (1.5) :
$$
\begin{array}{l}
W=\frac{1}{2} m_0 \chi_0^2 +\frac{1}{2} m_1 \chi_1^2 +
\frac{1}{2} m_2 \chi_2^2 \\
+m_3 \chi_1 \chi_2 
+ \frac{1}{3} \lambda(\chi_0^3 +\chi_1^3+\chi_2^3 +6 \chi_0\chi_1\chi_2).
\end{array}
\eqno(3.21)
$$

This will throw new light on 
the formula (1.5).

\section{Summary}

The charged lepton mass formula is not supernatural
 beings, so that the investigation should be done along
 the conventional mass matrix approach based on the
 field theory.

    A promising possibility has come up: 
The charged lepton mass formula (1.5) and the
 tribimaximal mixing (2.1) are related each other
 and both can be understood from S$_3$ or A$_4$ .

   Our next step is to attack the quark masses 
and mixing by applying the present approach.

\section*{Acknowledgements}

The author would like to thank participants to the workshop
NMM2006 for their valuable and helpful discussions.
He also thank E.~Ma for informing his recent paper\cite{Ma0612}
prior to putting it on the hep-ph arXive.
This work is supported by the Grant-in-Aid for
Scientific Research, Ministry of Education, Science and 
Culture, Japan (No.18540284).



\begin{thebibliography}{0}
\bibitem{Koide_Cab} Y.~Koide, preprint at Faculty of Engineering
Science, Osaka University, (1971), unpublised.
%
\bibitem{Hiroshima}  T.~Hayashi, Y.~Koide, S.~Ogawa, 
Progr.~Theor.~Phys. {\bf 39} (1968) 1372. 
%
\bibitem{Harari} H.~Harari, H.~Haut and J.~Weyers, Phys.~Lett. 
{\bf 78B} (1978) 459.
%
\bibitem{Koidemass} Y.~Koide, Lett.~Nuovo Cimento {\bf 34} (1982) 201;
Phys.~Lett. {\bf B120} (19836) 161;
Phys.~Rev. {\bf D28} (1983) 252.
%
%
\bibitem{Koide90} Y.~Koide, Mod.~Phys.~Lett. {\bf A5} (1990) 2319.
%
\bibitem{S3} S.~Pakvasa and H.~Sugawara, Phys.~Lett. {\bf B73} (1978) 61; 
H.~Harari, H.~Haut and J.~Weyers, Phys.~Lett. {\bf B78} (1978) 459; 
E.~Derman, Phys.~Rev. {\bf D19} (1979) 317;
D.~Wyler, Phys.~Rev. {\bf D19} (1979) 330.
%
\bibitem{Koide99} Y.~Koide, Phys.~Rev. {\bf D60} (1999) 077301.
%
%
\bibitem{tribi} 
S.~Pakvasa and H.~Sugawara, Phys.~Lett. {\bf B82} (1979) 105;
Y.~Yamanaka, H.~Sugawara and S.~Pakvasa, Phys.~Rev. {\bf D25} (1982) 1895;
P.~F.~Harrison, D.~H.~Perkins and W.~G.~Scott,
 Phys.~Lett. {\bf B458} (1999) 79;
 Phys.~Lett. {\bf B530} (2002) 167;
Z.~Z.~Xing, Phys.~Lett. {\bf B533} (2002) 85;
P.~F.~Harrison and W.~G.~Scott,
 Phys.~Lett. {\bf B535} (2003) 163;
Phys.~Lett. {\bf B557} (2003) 76;
E.~Ma, Phys.~Rev.~Lett. {\bf 90} (2003) 221802;
C.~I.~Low and R.~R.~Volkas, Phys.~Rev. {\bf D68} (2003) 033007;
X.-G.~He and A.~Zee,  Phys.~Lett. {\bf B560} (2003) 87.
%
\bibitem{Koide0603} Y.~Koide, arXive: hep-ph/0605074,
to be published in J.~Phys. G (2007).
%
\bibitem{UnivSeesaw} The seesaw mechanism for charged particles is known
as the ``universal seesaw mechanism":
Z.~G.~Berezhiani, Phys.~Lett.~{\bf 129B} (1983) 99;
Phys.~Lett.~{\bf 150B} (1985) 177;
D.~Chang and R.~N.~Mohapatra, Phys.~Rev.~Lett.~{\bf 58} (1987) 1600; 
A.~Davidson and K.~C.~Wali, Phys.~Rev.~Lett.~{\bf 59} (1987) 393;
S.~Rajpoot, Mod.~Phys.~Lett. {\bf A2} (1987) 307; 
Phys.~Lett.~{\bf 191B} (1987) 122; Phys.~Rev.~{\bf D36} (1987) 1479;
K.~S.~Babu and R.~N.~Mohapatra, Phys.~Rev.~Lett.~{\bf 62} (1989) 1079; 
Phys.~Rev. {\bf D41} (1990) 1286; 
S.~Ranfone, Phys.~Rev.~{\bf D42} (1990) 3819; 
A.~Davidson, S.~Ranfone and K.~C.~Wali, 
Phys.~Rev.~{\bf D41} (1990) 208; 
I.~Sogami and T.~Shinohara, Prog.~Theor.~Phys.~{\bf 86} (1991) 1031;
Phys.~Rev. {\bf D47} (1993) 2905; 
Z.~G.~Berezhiani and R.~Rattazzi, Phys.~Lett.~{\bf B279} (1992) 124;
P.~Cho, Phys.~Rev. {\bf D48} (1993) 5331; 
A.~Davidson, L.~Michel, M.~L.~Sage and  K.~C.~Wali, 
Phys.~Rev.~{\bf D49} (1994) 1378; 
W.~A.~Ponce, A.~Zepeda and R.~G.~Lozano, 
Phys.~Rev.~{\bf D49} (1994) 4954.
%
\bibitem{Koide06} Y.~Koide, Phys.~Rev. {\bf D73} (2006) 057901.
Also, see Y.~Koide, Phys.~Rev. {\bf D60} (1999) 077301.
%
\bibitem{Koide0605} Y.~Koide, arXive: hep-ph/0612058, to be
published in Euro.~Phys.~J C (2007).
%
\bibitem{atm} 
D.~G.~Michael {\it et al.}  MINOS Collaboration, Phys.~Rev.~Lett.
{\bf 97} (2006) 191801.
Also see, J.~Hosaka {\it et al.} the Super-Kamiokande Collaboration,
arXiv: hep-ex/0604011.
%
\bibitem{solar} 
B.~Aharmim {\it et al.} SNO Collaboration,
Phys.~Rev. {\bf C72} (2005) 055502;
T.~Araki  {\it et al.} KamLAND Collaboration, 
Phys.~Rev.~Lett. {\bf 94} (2005) 081801.
%
\bibitem{23sym}
T.~Fukuyama and H.~Nishiura,
hep-ph/9702253, in Proceedings of {\it the International 
Workshop on Masses and 
Mixings of Quarks and Leptons},  Shizuoka, Japan, 1997, 
edited by Y.~Koide (World Scientific, Singapore, 1998), p. 252; \\
E.~Ma and  M.~Raidal, 
{\it Phys.~Rev.~Lett.} {\bf 87} (2001) 011802 [hep-ph/0102255]; \\
C.~S.~Lam, 
{\it Phys.~Lett.} {\bf B507} (2001) 214 [hep-ph/0104116]; \\
K.~R.~S.~Balaji, W.~Grimus and T.~Schwetz, 
{\it Phys.~Lett.} {\bf B508} (2001) 301 [hep-ph/0104035]; \\
W.~Grimus and L.~Lavoura, 
{\it Acta Phys.~Pol.} {\bf B32} (2001) 3719 [hep-ph/0110041].
%
\bibitem{Brannen} C.~Brannen, 
http://brannenworks.com/MASSES2.pdf, (2006).
%
\bibitem{Rosen} G.~Rosen, Mod.~Phys.~Lett. {\bf A22}, 283 (2007).
%
\bibitem{Frogatt} C.~Frogatt and H.~B.~Nielsen, Nucl.~Phys. 
{\bf B147} (1979) 277.
%
%
\bibitem{evol_KoideF}
N.~Li and B.~Q.~Ma, Phys.~Rev. {\bf D73} (2006) 013009;
Z.~Z.~Xing, H.~Zhang, Phys.~Lett. {\bf B635} (2006) 107.
%
\bibitem{Ma06}
E.~Ma, Phys.~Rev. {\bf D73} (2006) 057304.
%
\bibitem{Koide-A4} Y.~Koide, arXive: hep-ph/0701018.
%
\bibitem{Ma-priv}
E.~Ma, private communication (2006).
%
\bibitem{Ma0612}  E.~Ma, arXive: hep-ph/0612022.
%
%

%

\end{thebibliography}
\end{document}